\documentclass[prl,aps,showpacs,a4paper,superscriptaddress,twocolumn,floatfix]{revtex4}
\usepackage{latexsym}
\usepackage{amsmath, mathrsfs}
\usepackage[latin1]{inputenc}
\usepackage[dvips]{graphicx}

\newcommand{\xbj}{{x}}
\newcommand{\qsa}{{Q_\mathrm{s,A}}}
\newcommand{\qssa}{{Q^2_\mathrm{s,A}}}

\newcommand{\qssb}{{Q^2_\mathrm{s,B}}}

\newcommand{\qssp}{{Q^2_\mathrm{s,p}}}
\newcommand{\qs}{{Q_\mathrm{s}}}
\newcommand{\qss}{{Q^2_\mathrm{s}}}
\newcommand{\lqcd}{\Lambda_{\mathrm{QCD}}}
\newcommand{\as}{{\alpha_{\mathrm{s}}}}
\newcommand{\rt}{{\mathbf{r}_\perp}}

\newcommand{\bt}{{\mathbf{b}_\perp}}
\newcommand{\Deltat}{{\boldsymbol{\Delta}_\perp}}

\newcommand{\nc}{{N_\mathrm{c}}}

\newcommand{\gev}{\textrm{ GeV}}
\newcommand{\fm}{\textrm{ fm}}
\newcommand{\mb}{\textrm{ mb}}
\newcommand{\ra}{R_A}
\newcommand{\rp}{R_p}

\newcommand{\Aavg}[1]{\left\langle #1 \right\rangle_\textrm{N}}

\newcommand{\nr}[1]{(\ref{#1})} 
\newcommand{\ud}{\, \mathrm{d}}
\newcommand{\fig}{Fig.~}
\newcommand{\eq}{Eq.~}

\newcommand{\eqs}{Eqs.~}

\newcommand{\sigmap}{{ \sigma^\textrm{p}_\textrm{dip} }}
\newcommand{\sigmadip}{{ \sigma_\textrm{dip} }}
\newcommand{\sigmadipsqr}{{ \sigma^2_\textrm{dip} }}
\newcommand{\dsigmap}{{\frac{\ud \sigma^\textrm{p}_\textrm{dip}}{\ud^2 \bt}}}
\newcommand{\dsigmaa}{{\frac{\ud \sigma^A_\textrm{dip}}{\ud^2 \bt}}}

\newcommand{\dsigma}{{\frac{\ud \sigma_\textrm{dip}}{\ud^2 \bt}}}

\begin{document}

\title{
Nuclear enhancement of universal dynamics of high parton densities}

\preprint{arXiv:0705.3047 [hep-ph]}

\author{H. Kowalski}
\affiliation{Deutsches Elektronen-Synchrotron DESY, 22607 Hamburg, Germany}
\author{T. Lappi}
\affiliation{Physics Department, Brookhaven
National Laboratory, Upton, NY 11973, USA}
\author{R. Venugopalan}
\affiliation{Physics Department, Brookhaven
National Laboratory, Upton, NY 11973, USA}

\begin{abstract}
We show that the enhancement of the saturation scale in large nuclei relative to the proton is significantly influenced by the effects of quantum evolution and the impact parameter dependence of dipole cross sections in high energy QCD. We 
demonstrate that there is a strong $A$ dependence in diffractive deeply inelastic scatteringand discuss its sensitivity to the measurement of the recoil nucleus.
\end{abstract}


\maketitle

The properties of hadronic and nuclear wave functions at high energies are of great importance in understanding 
multi-particle production in QCD. Especially intriguing is the possibility that the small Feynman $x$ components 
of these wavefunctions demonstrate universal behavior that is insensitive to the details of hadron or nuclear structure 
in the (large $x$) fragmentation region. 

The specific nature of universal small $x$ dynamics in QCD follows from the strong enhancement
of gluon bremsstrahlung at small $x$
leading to a rapid growth of the occupation number of a transverse
momentum mode $k_\perp$ in the hadron or nuclear wavefunction.  
However, it can maximally be of order $1/\as$ (where $\as$ is the QCD coupling constant)
because of non-linear multi--parton effects such as recombination and screening which deplete
the gluon density at small $x$~\cite{Gribov:1984tu}.
In particular, the occupation number is maximal for modes with
$k_\perp \lesssim \qs$, where $\qs(x)$, appropriately called the saturation scale, is a scale
generated by the multi-parton dynamics. For a probe with transverse resolution $1/Q^2$, this
scale is manifest in a universal scaling form of observables as a function of $Q/\qs$ in a wide
kinematical range in $x$ and $Q^2$.

In addition to the strong $x$ dependence generated by gluon bremsstrahlung, the saturation scale $\qs$ has a strong $A$ dependence because 
of the Lorentz contraction, in the probe rest frame, of the nuclear parton density. 
For large enough $A$ and small enough $x$, the saturation scale is larger than $\lqcd$, 
the fundamental soft scale of QCD. 
In this letter, we will discuss the $A$ and $x$ dependence of the 
saturation scale 
and some of its ramifications for hard diffraction in nuclei.

A saturation scale arises naturally in the Color Glass 
Condensate (CGC)~\cite{Iancu:2003xm}
description of universal properties of hadron and
 nuclear wavefunctions at small $x$. The CGC, when applied to Deeply Inelastic 
Scattering (DIS), 
results~\cite{McLerran:1998nk,Venugopalan:1999wu}, at leading order in 
$\as$, in the dipole picture of 
DIS~\cite{Mueller:1989st},
where the inclusive virtual photon hadron cross section is
\begin{equation}\label{eq:sigmatot}
\sigma^{\gamma^*p}_{L,T}
= \int\! \ud^2 \rt \int_0^1 \! \ud z \left| \Psi^{\gamma^*}_{L,T}
\right|^2 
\int \! \ud^2 \bt \dsigmap    .
\end{equation}
Here $\left| \Psi_{L,T}^{\gamma^*}(\rt,z,Q) \right|^2$
represents the probability for a  virtual photon to produce a quark--anti-quark pair of size $r = |\rt|$ and $\dsigmap(\rt,\xbj,\bt)$ denotes the \emph{dipole cross section} for this pair to scatter off the target at an impact parameter $\bt$. The former is well known from QED, while the latter represents the dynamics of QCD scattering at small $x$. A simple saturation model 
(known as the GBW model~\cite{Golec-Biernat:1998js}) 
of the dipole cross section, parametrized as 
$\dsigmap = 2 ( 1 - e^{ - r^2 \qssp(x)/4})$ where 
$\qssp (x) = (x_0/x)^\lambda \gev^2$, gives a good qualitative fit to the HERA inclusive cross section data for 
$x_0 = 3\cdot 10^{-4}$ and $\lambda = 0.288$. 
However, the model
does not contain the bremsstrahlung limit of perturbative QCD (pQCD)
that applies to small dipoles of size $r \ll 1/\qs(x)$. 

In the classical effective theory of the 
CGC, to leading logarithmic accuracy, 
one can derive the dipole cross section~\cite{Venugopalan:1999wu} 
containing the right small $r$ limit.
This dipole cross section can be 
represented (see however \cite{GolecBiernat:2003ym})
\begin{equation}
\dsigmap
 = 2\,\left[ 1 - \exp\left(- r^2  F(\xbj,r) T_p(\bt)\right) 
\right],
\label{eq:BEKW}
\end{equation}
where $T_p(\bt)$ is the impact parameter profile function in the proton, normalized as 
$\int d^2 \bt \,T_p(\bt) = 1$ and $F$ is proportional to the 
DGLAP evolved gluon  distribution~\cite{Bartels:2002cj}
\begin{equation}
F(\xbj,r^2) = \pi^2 \as\left(\mu_0^2 + 4/r^2 \right) 
\xbj g\left(\xbj,\mu_0^2 + 4/r^2 \right)/(2 \nc).
\label{eq:BEKW_F}
\end{equation}
The dipole cross section in \eq\nr{eq:BEKW} was 
implemented in the impact parameter saturation model (IPsat)~\cite{Kowalski:2003hm} where the parameters are fit to 
reproduce the HERA data on the inclusive structure function $F_2$. 

In general, the dipole cross section can range from $0$ in the
$r \to 0$ color 
transparency limit to $2$, the maximal unitarity bound.
The saturation scale $\qs$ characterizes the qualitative change
between these regimes;
we shall here define $\qs$ as the solution 
of $\dsigma(\xbj , r^2 = 1/\qss(\xbj,\bt)) = 2(1-e^{-1/4})$~\footnote{
Our definition is equivalent to the 
saturation scale in the GBW model for a Gaussian dipole cross section. Note the difference with the convention in 
Ref.~\cite{Kowalski:2003hm}.}.

The IPsat dipole cross section in \eq\nr{eq:BEKW} is applicable
when leading logarithms in $Q^2$ dominate over leading logarithms in $x$. 
At very small $x$, quantum evolution in the CGC~\cite{Iancu:2003xm} describing 
both the bremsstrahlung limit of linear small $x$ evolution as well as nonlinear RG
evolution at high parton densities, combined with a realistic $b$-dependence, is better captured in the 
bCGC model~\cite{Iancu:2003ge,Kowalski:2006hc}.
Both the IPsat model and 
the bCGC model provide excellent fits 
to a wide range of HERA data for $x \leq 0.01$~\cite{Kowalski:2006hc,Forshaw:2006np}. 
We will now discuss the possibility that DIS off nuclei can distinguish respectively
 between these  ``classical CGC'' and ``quantum CGC'' motivated models. 

A straightforward generalization of the dipole formalism to nuclei
is to introduce the coordinates of the 
individual nucleons $\{\bt_i\}$. 
One obtains in the IPsat model, 
\begin{equation}\label{eq:defsigmaa}
\dsigmaa
= 2 \left[1- 
e^{- r^2
F(\xbj,r) \sum_{i=1}^A T_p(\bt-\bt_i)
}
\right],
\end{equation}
where $F$ is defined in \eq\nr{eq:BEKW_F}. The positions of the nucleons
$\left\{\bt_i\right\}$
are distributed 
according to the Woods-Saxon distribution $T_A(\bt_i)$.
We denote the average of an observable $\mathcal{O}$
 over $\left\{\bt_i\right\}$ by
$\Aavg{\mathcal{O}} \equiv \int \prod_{i=1}^A \ud^2 \bt_i 
T_A(\bt_i) \mathcal{O}(\left\{\bt_i\right\})
$. 
The average differential dipole cross section is well approximated by\cite{Kowalski:2003hm} 
\begin{equation}\label{eq:glauberdsigma}
\Aavg{\dsigmaa} 
\approx 2\left[1-\left(1-\frac{T_A(\bt)}{2}\sigmap \right)^A\right]
\end{equation}
where, for large $A$, the expression in parenthesis can be replaced by
 $\exp\left(-\frac{A T_A(\bt)}{2}\sigmap \right)$~\cite{Gotsman:1999vt}. 
All parameters of the model come from either fits of the model to 
$ep$-data or from the Woods-Saxon distributions; no additional parameters are
 introduced for $eA$ collisions. The same exercise is repeated 
for the bCGC model.

\begin{figure}
\includegraphics[width=0.4\textwidth]{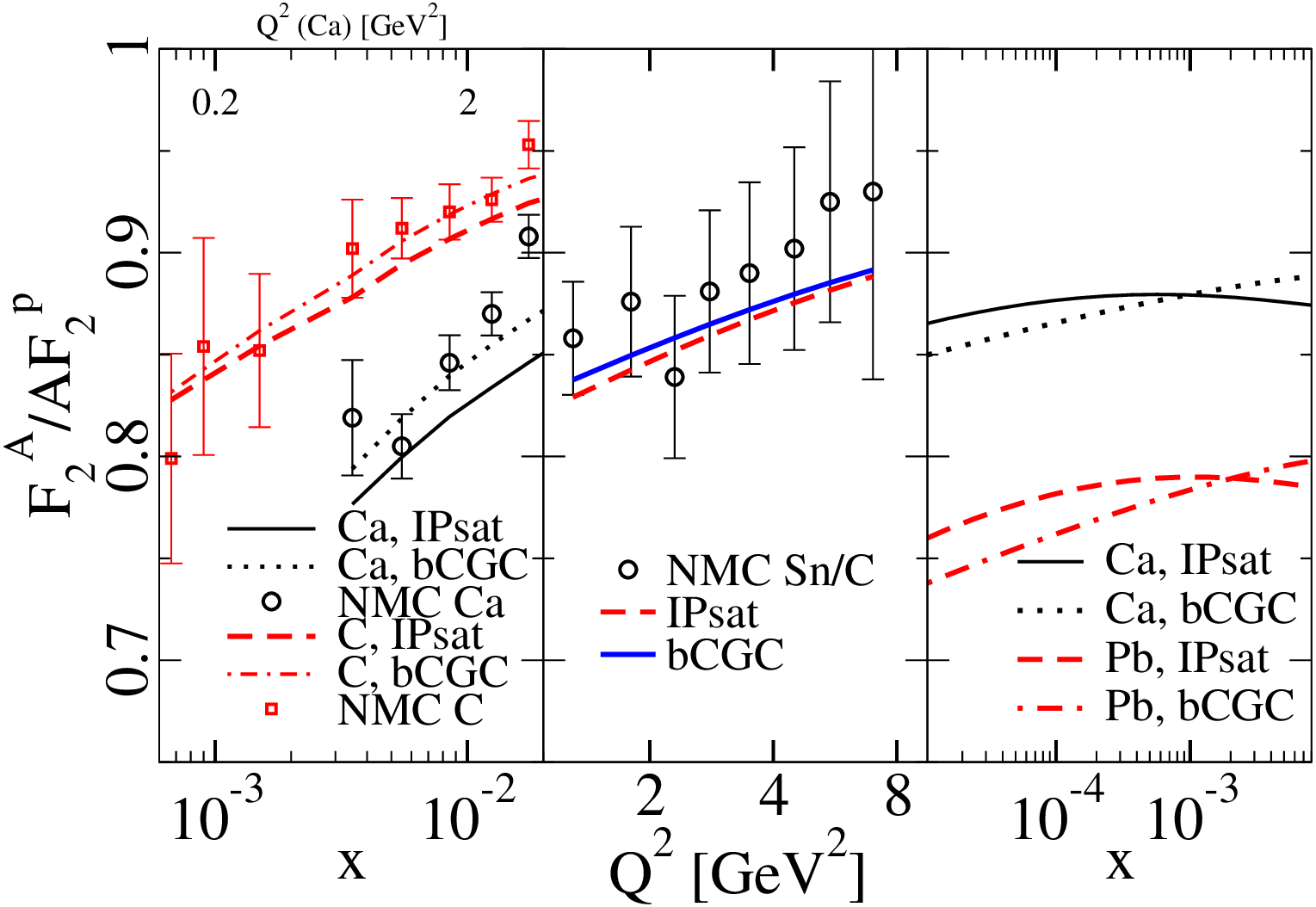}
\caption{
Left: Predictions for shadowing compared to NMC data.
Center: predictions for $12F_2^\textrm{Sn}/118F_2^\textrm{C}$ compared
to NMC data at $\xbj = 0.0125$.
Right: Likewise for $Q^2 = 5 \gev^2$ as a function of $\xbj$.
}
\label{fig:shad}
\end{figure}

In \fig\ref{fig:shad} (left), we compare the prediction of the IPsat and bCGC models
 with the experimental data~\footnote{
Data available from the E665 collaboration differ from the NMC data for 
a similar kinematic range.} on nuclear DIS from the NMC
 collaboration~\cite{Amaudruz:1995tq}). 
Figure~\ref{fig:shad} (right) shows that the $\xbj$ dependence of shadowing for 
fixed $Q^2$ in the IPsat model is very flat.
This is because the best fit to $ep$-data in DGLAP-based dipole 
models~\cite{Bartels:2002cj,Kowalski:2003hm} is given by a 
very weak $\xbj$-dependence at the initial scale
$\mu_0^2$. A stronger $\xbj$-dependence also for large dipoles, such as in the 
in the GBW or bCGC models, gives a stronger $\xbj$-dependence of
shadowing at fixed $Q^2$. As shown in \fig\ref{fig:shad} (center),
 both the IPsat and bCGC models predict strong
$Q^2$-dependence (at fixed $x$) for shadowing. 
It is this latter effect which is primarily responsible
for the shadowing effect seen in the  NMC data. 
Precision measurements of $F_2^A/AF_2^p$ would shed more light on the
relative importance of $Q^2$ and $\xbj$ evolution in this regime.

We now turn to a discussion of the $A$ and $x$ dependence of the saturation scale. In a simple GBW type model, inserting a $\theta$-function impact parameter dependence into \eq\nr{eq:glauberdsigma} yields the estimate
$\qssa \approx A^{1/3}\frac{\rp^2 A^{2/3}}{\ra^2}\qssp
\approx 0.26 A^{1/3} \qssp$ for $ 2\pi \rp^2 \approx 20\mb $ and $\ra\approx 1.1\,A^{1/3}\fm$. The smallness of $\qssa/\qssp$, due to the 
constant factor $\sim 0.26$
has sometimes been interpreted~\cite{Freund:2002ux,Kowalski:2003hm,Kopeliovich:2006bm} as a weak nuclear enhancement of $\qs$.
We will argue here that detailed considerations of QCD evolution 
and the $b$-dependence of the dipole cross section  
result in a significantly larger nuclear enhancement
of $\qs$.

The effect of QCD evolution on $\qsa$ in the IPsat nuclear dipole cross section is 
from the 
DGLAP-like growth of the gluon distribution. The increase in the gluon density  with 
increasing $Q^2$ 
and decreasing (dominant) dipole radius $r$ causes 
$\qs$ grow even faster as a function of $A$. This is seen qualitatively for two
 different nuclei, $A$ and $B$ (with $A>B$), in a ``smooth nucleus'' approximation of \eq\nr{eq:defsigmaa} whereby 
$\sum_{i=1}^A T_p(\bt-\bt_i)$ is replaced by $A\,T_A(\bt)$. We 
obtain
\begin{equation}
\frac{\qssa}{\qssb} = 
\frac{A}{B}\frac{T_A(\bt)}{T_B(\bt)}
\frac{F(\xbj,\qssa)}{F(\xbj,\qssb)} 
\sim 
\frac{A^{1/3}}{B^{1/3}} \frac{F(\xbj,\qssa)}{F(\xbj,\qssb)}.
\end{equation}
The scaling violations in $F$ imply that,
as observed in Refs.~\cite{Kowalski:2003hm,Armesto:2004ud},
the growth of $\qs$ is faster than $A^{1/3}$.
Also, because the increase  of $F$ with $Q^2$ is faster for smaller $x$, the 
$A$-dependence of $\qs$ is stronger for higher energies.
In contrast, the dipole cross section in the bCGC model depends only on the 
``geometrical scaling" combination~\footnote{
See, however
Ref.~\cite{Albacete:2007yr} on
``pre-asymptotic" violations of geometrical scaling. }
$r \qs(x)$ without DGLAP scaling violations and therefore does not have this 
particular nuclear enhancement~\footnote{Another 
interesting possibility that running coupling $\log \xbj$-evolution 
results in a depletion of the $A$-dependence of $\qs$~\cite{Mueller:2003bz}. 
}.
Precise extraction of the $A$ dependence of $\qs$ will play an important role in 
distinguishing between ``classical'' and ``quantum'' evolution in the CGC.

\begin{figure}
\includegraphics[width=0.34\textwidth]{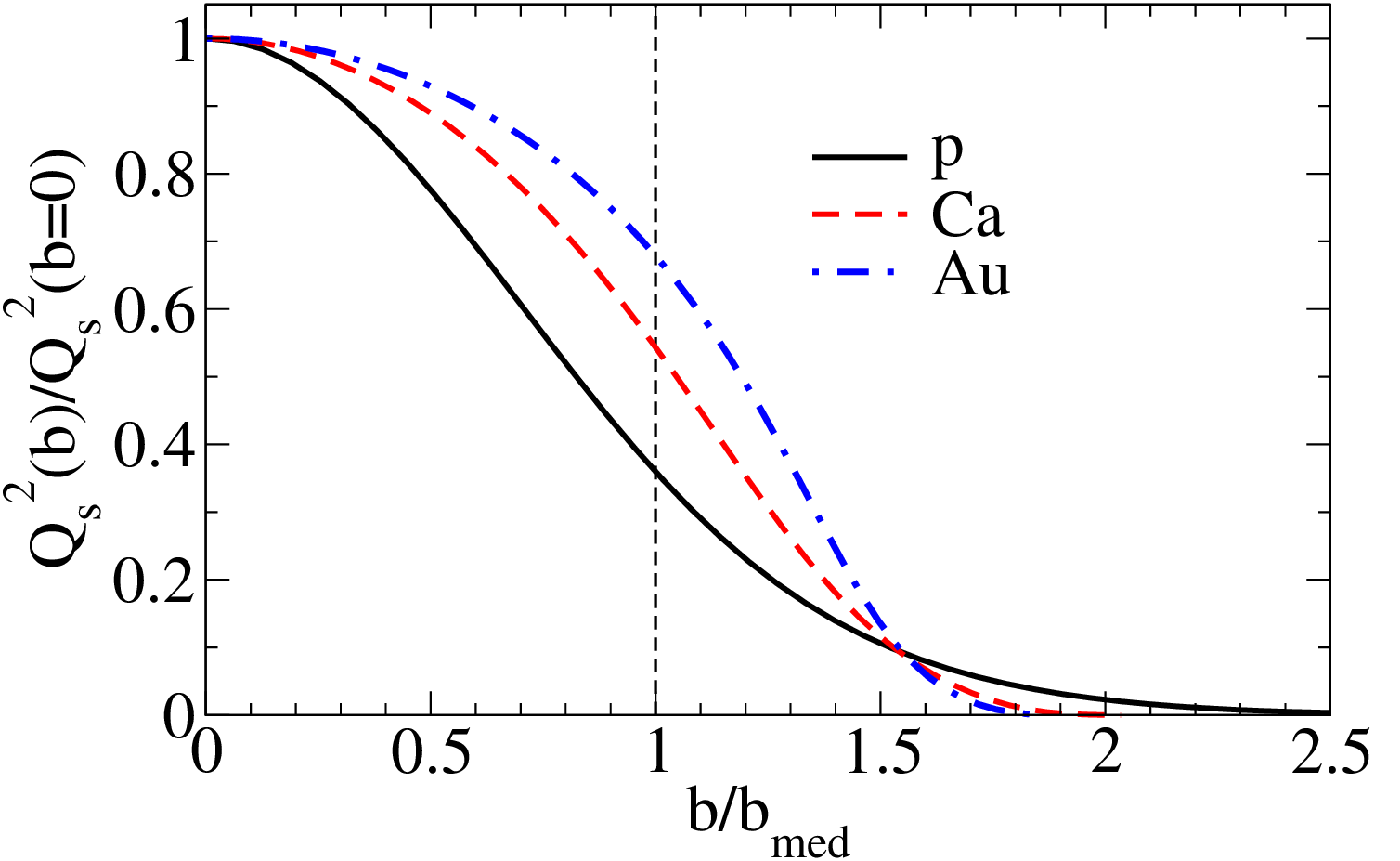}
\caption{
Impact parameter dependence of the saturation scale for p, Ca and Au 
at $\xbj = 0.001$ and $Q^2 = 1 \gev^2$. Details in text.
}
\label{fig:Qsabt}
\end{figure}

A careful evaluation shows that because the density 
 profile in a nucleus is more uniform than that of the proton, the 
saturation scales  in nuclei decrease more slowly with 
$b$ than in the proton. The dependence of the saturation scale on the impact 
parameter is plotted in \fig\ref{fig:Qsabt}. The saturation scale 
in gold nuclei at the median impact parameter for the total cross section
$b_\textrm{med.}$ is about 70\% of the value at 
$b=0$; in contrast, $\qssp(b_\textrm{med.})$ is  only $\sim 35$\% of the value at $b=0$. 

\begin{figure}
\hfill
\includegraphics[width=0.2401\textwidth]{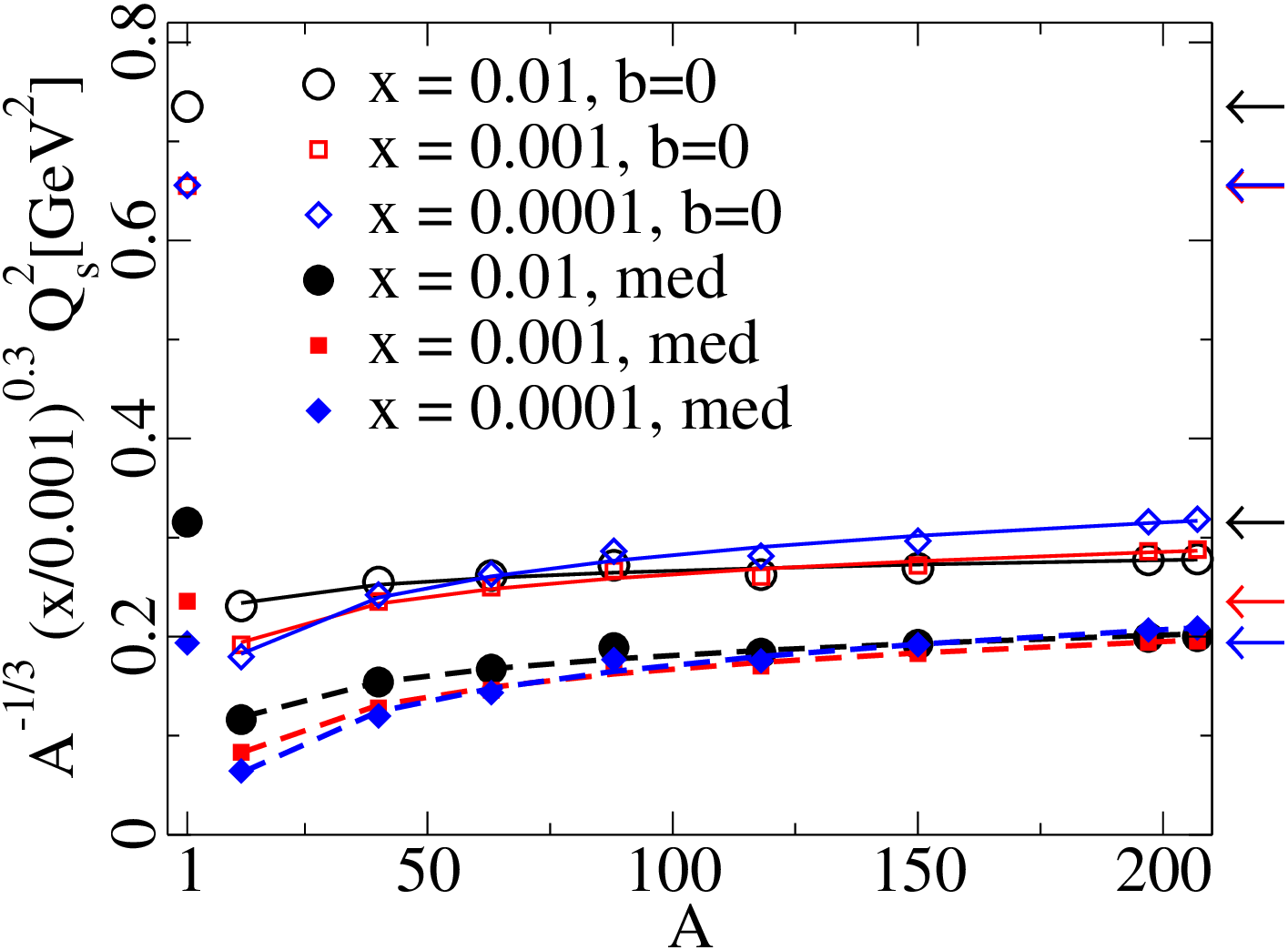}
\hfill
\includegraphics[width=0.2200\textwidth]{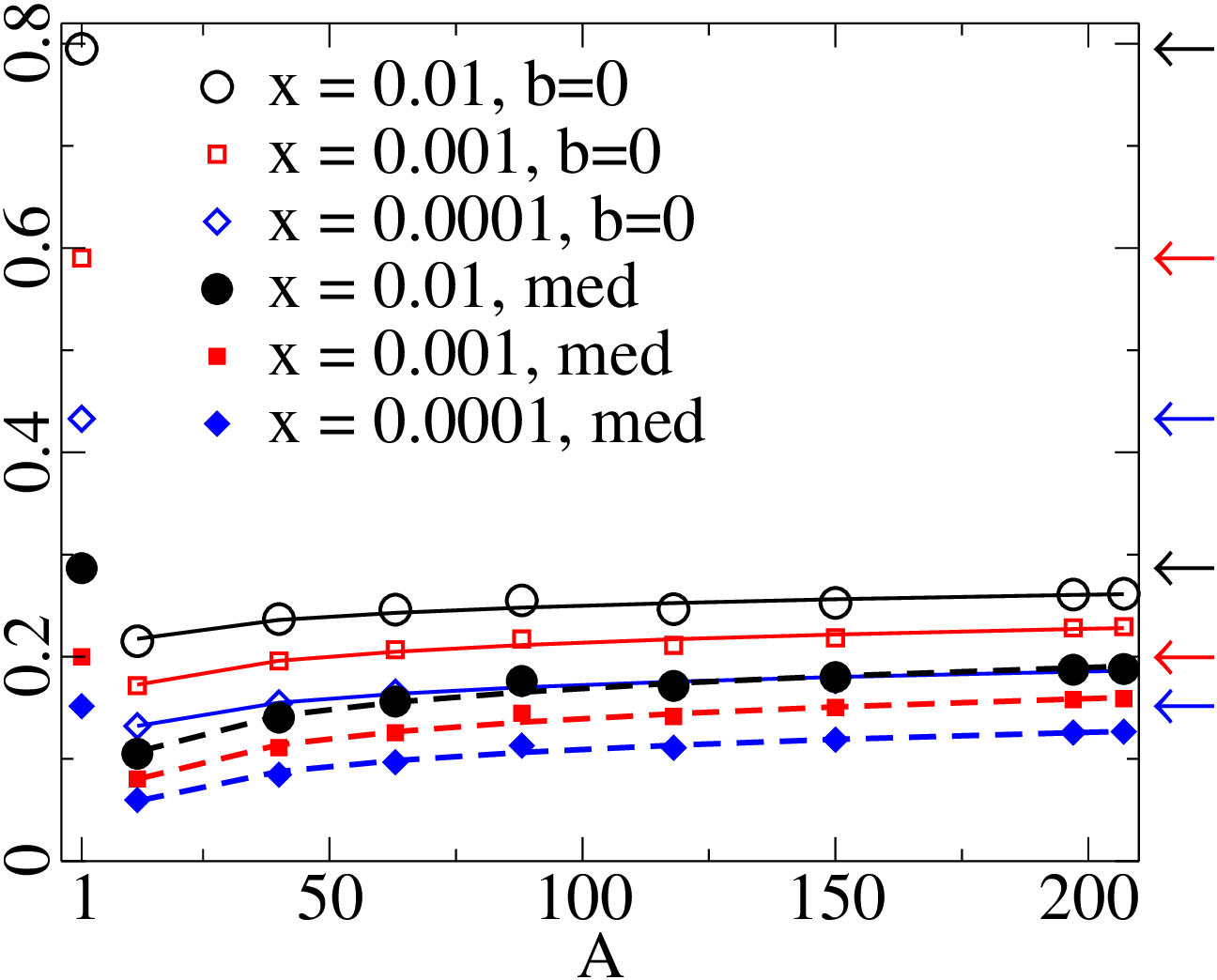}
\hfill
\caption{
Saturation scale at $b = 0$ (open symbols) and $b=b_\textrm{med.}$ (filled symbols) as a function of $A$ for different 
$\xbj$. The saturation scale for the proton is shown at $A=1$
and by the arrows on the right.
Left: IPsat model. Right: bCGC model.
}
\label{fig:Qscombination}
\end{figure}

The $A$ dependence of the saturation scale for various $x$
is shown in \fig\ref{fig:Qscombination}, for the IPsat model on the left
and the bCGC model on the right.
Note that in the IPsat model, at small $x$, $\qssa(b_\textrm{med.})$ for gold nuclei
is nearly identical to $A^{1/3}$ times the value for the proton. 
The corresponding enhancement 
for $b=0$ is significantly smaller as anticipated. 
The nuclear enhancement in the bCGC model is nearly as large, 
showing that it  
owes, for the kinematic range studied, 
much more to the relative impact parameter profiles 
 (see \fig\ref{fig:Qsabt}) than to differences in QCD evolution. 
Nevertheless, the stronger $A$ dependence of 
$\qssa(b_\textrm{med.})$ in the IPsat model relative to the 
bCGC model, especially at the smallest $x$ values, 
clearly illustrates the differences in quantum evolution between the models.
The factor of $200^{1/3} \approx 6$ 
gives a huge ``oomph'' in the parton density of a nucleus relative to that of a proton;
 one requires a center of mass energy 
\emph{$\sim 14$ times larger in an e+p collider relative to an e+Au collider} to obtain
 the same $\qssa(b_\textrm{med.})(x)$.

We will now focus on some 
interesting qualitative features of hard diffraction off nuclei
(see also \cite{NikStrik}). For simplicity, we will consider only the IPsat model here. 
The contribution of  $q\bar{q}$ dipoles~\footnote{
The contribution of higher
Fock states will be addressed in future work.}
to the inclusive diffractive cross section can be expressed as
\begin{equation}\label{eq:sigmadiff}
\frac{\ud \sigma^D_{L,T}}{\ud t} =
\frac{1}{16\pi}
\int \! \ud^2 \rt 
\ud z
\left| \Psi^{\gamma^*}_{L,T}
\right|^2
\sigmadipsqr(\xbj,r,\Deltat),
\end{equation}
where $t = -\Deltat^2$ and $\sigmadip(\xbj,r,\Deltat)$
is the Fourier transform of the dipole cross section with respect to $\bt$.  
The total diffractive cross section, obtained by integrating \eq\nr{eq:sigmadiff} over $t$, reads
\footnote{
A comparison of  \eqs\nr{eq:sigmatot} and \nr{eq:sigmadifftot} makes explicit the
unitarity limit  
$\sigma^D_{L,T} \leq \sigma_{L,T}/2$, which is saturated by a black disk,
$\dsigma(\bt) = 2\theta(R-b),$ leading to the 
$t$-distribution
$\sim 4 |J_1(\sqrt{-t}R)|^2/(-t R^2) \approx 1+tR^2/4+\mathcal{O}(t^2)$. 
In contrast, an exponential $t$ distribution 
$\sim e^{Bt}$ (see e.g.~\cite{Kugeratski:2005ck})
 corresponds to a Gaussian in $b$, where the unitarity limit,  saturated by 
$\dsigma = 2 \, e^{-b^2/(2 B)}$, is $\sigma^D_{L,T} \leq \sigma_{L,T}/4$.
Specifying the $\bt$ dependence of the dipole cross 
section simultaneously fixes both 
the normalization $\sigma_0$ of the cross section and the diffractive slope 
$B = \ud \ln \sigma^D/\ud t|_{t=0}$. ($\sigma_0/B = 8 \pi$ and $\sigma_0/B = 4 \pi$ for black disk and 
Gaussian respectively.)
}
\begin{equation}\label{eq:sigmadifftot}
\sigma^D_{L,T} = \frac{1}{4}
\int\! \ud^2 \rt 
\ud z
\left| \Psi^{\gamma^*}_{L,T}
\right|^2
\int \ud^2\bt
\left( \dsigma\right)^2.
\end{equation}

The diffractive slope at $t=0$ depends on the size of the system.
For small $t\sim - 1/\ra^2$ one expects a very steep 
$t$-dependence $\sim \exp\{D t \ra^2\}$ (with $D \sim 1$). In our picture of the 
 nucleus as a ``lumpy'' collection of partially overlapping
nucleons (\eqs\nr{eq:defsigmaa} and \nr{eq:glauberdsigma}), an interesting question 
is whether this lumpiness shows up as a 
proton-like tail $\sim \exp\{D' t\rp^2)\}$ of the $t$-distribution. 

\begin{figure}
\includegraphics[width=0.38\textwidth]{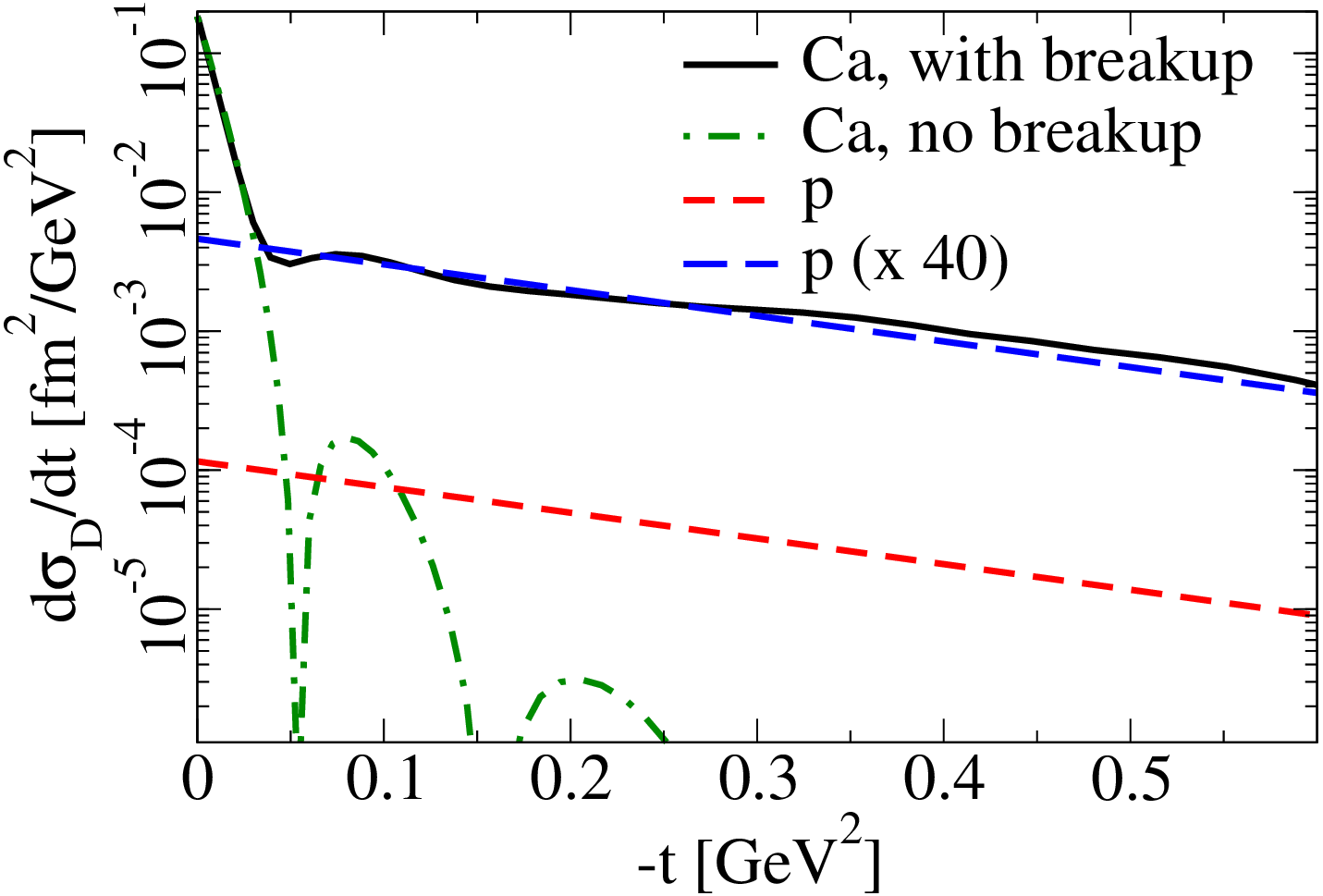}
\caption{
The $t$-dependence of the calcium and the proton dipole cross sections.
 The ``breakup'' curve for calcium 
is computed using $\frac{1}{16 \pi} \Aavg{|\sigmadip(\xbj,r,\Deltat)|^2}$
 for $r = 0.2\fm$ and $\xbj  = 0.001$ and ``no breakup'' curve using 
 $\frac{1}{16 \pi} \Aavg{\sigmadip(\xbj,r,\Deltat)}^2$. 
}
\label{fig:tdep}
\end{figure}

If one requires that the nucleus stays completely intact, the 
average $\Aavg{\cdot}$ must be performed at the amplitude level, and 
$\ud\sigma^D/\ud t$ falls off very rapidly as
$\sim \exp\{D t \ra^2\}$. Measuring the intact recoil nucleus at such a
 small $t$ experimentally 
at a future electron ion collider~\cite{Deshpande:2005wd}
is challenging. Considerable physical insight into the diffractive process 
can be obtained in events where the nucleus breaks up into color neutral 
constituents without filling 
the rapidity gap between the $q\bar{q}$ dipole and the nuclear 
fragmentation region. Such events correspond to performing the average 
$\Aavg{\cdot}$ over the cross section~\cite{Kovchegov:1999kx},
\eq\nr{eq:sigmadifftot}, instead of 
the amplitude. The difference between the two averaging procedures can be significant 
with increasing values of $t$; the result for calcium nuclei is shown in 
\fig\ref{fig:tdep}. The $t$ dependence of the proton is shown as well; as expected, the 
``break up'' cross section for calcium converges to $A$ times the  proton
 cross section with increasing $t$. 

\begin{figure}
\includegraphics[width=0.4\textwidth]{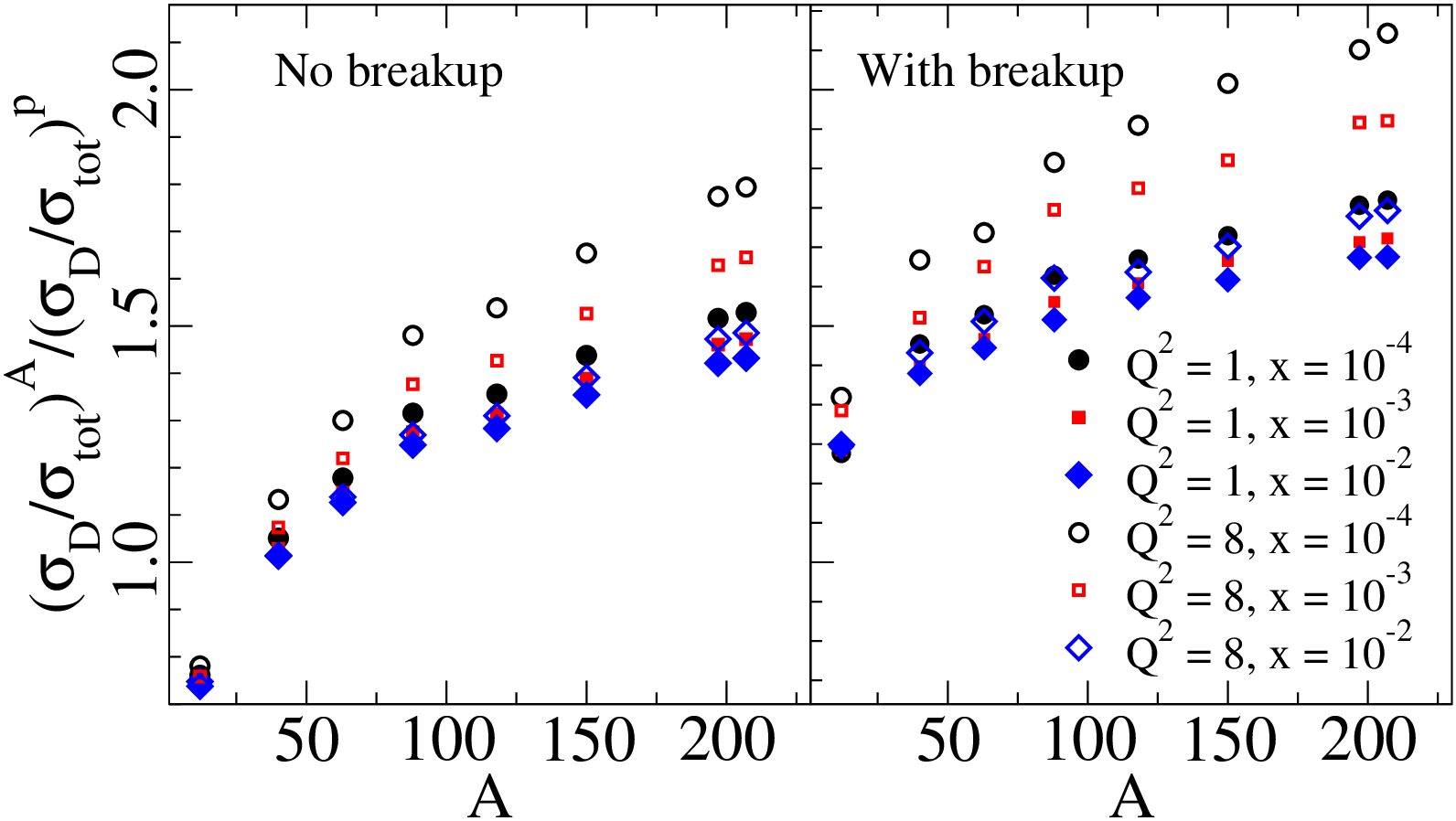}
\caption{
Ratio of the fraction of $q\bar{q}$ diffractive events in nuclei to the fraction in a  proton plotted  
versus atomic number $A$ for fixed $x$ and $Q^2$ values. 
}
\label{fig:sigmaratio}
\end{figure}

The difference between ``no break up'' and ``break up''  
integrated cross sections can be seen in 
\fig\ref{fig:sigmaratio} where we plot, as a function of 
$A$ for fixed $x$ and $Q^2$, the double ratio $R_\textrm{diff.}^A$, defined as 
the ratio $(\sigma^{D}_{T}  + \sigma^{D}_{L})/(\sigma^{\gamma^*}_{T}  +
\sigma^{\gamma^*}_{L})$ from \eqs\nr{eq:sigmatot} and \nr{eq:sigmadifftot} for a 
nucleus divided by the same ratio for a proton.
For light nuclei ($A < 40$), $R_\textrm{diff.}^A < 1$
 before going well above unity for large $A$. This is because
 the diffractive $q\bar{q}$ cross section is dominated by smaller impact parameters than the inclusive cross section; at small impact parameters, the 
matter density in a proton is larger than in light nuclei. For large nuclei, especially in the ``break up'' case, the nuclear enhancement of the 
fraction of diffractive events can be quite 
significant, up to a 100\% enhancement relative to the fraction for a  proton. 

\acknowledgments{We thank A. Caldwell, C. Marquet and L. Motyka for very useful 
discussions. This manuscript has been authorized under Contract No. DE-AC02-98CH10886 
with the U.S. Department of Energy.}

\end{document}